*Article*

# Trends on Computer Security: Cryptography, User Authentication, Denial of Service and Intrusion Detection

**Pablo Daniel Marcillo Lara** [†,‡], **Daniel Alejandro Maldonado-Ruiz** [†,‡], **Santiago Daniel Arrais Díaz** [†,‡], **Lorena Isabel Barona López** [†,‡] and **Ángel Leonardo Valdivieso Caraguay** [†,‡]

Departamento de Informática y Ciencias de la Computación (DICC), Escuela Politécnica Nacional, Quito, Ecuador

* Correspondence: angel.valdivieso@epn.edu.ec

† Current address: Departamento de Informática y Ciencias de la Computación (DICC), Escuela Politécnica Nacional, Ladrón de Guevara E11-25 y Andalucía, Edificio de Sistemas, 170525 Quito, Ecuador

‡ These authors contributed equally to this work.

**Abstract:** The new generation of security threats has been promoted by digital currencies and real-time applications, where all users develop new ways to communicate on the Internet. Security has evolved in the need of privacy and anonymity for all users and his portable devices. New technologies in every field prove that users need security features integrated into their communication applications, parallel systems for mobile devices, internet, and identity management. This review presents the key concepts of the main areas in computer security and how it has evolved in the last years. This work focuses on cryptography, user authentication, denial of service attacks, intrusion detection and firewalls.

---

## 1. Introduction

The increasing number of heterogeneous devices connected to Internet, running different OS/Apps and transporting private users information has notably promoted attackers to develop new varieties of security threats. Some of the security solutions proposed and adopted in the past are no longer viable in the last generation of devices. The computing capacity has been increased not only by the individual devices but also the possibility to access to distributed resources that provide higher capabilities to users. In this context, it is important to have a clear view of the latest tendencies on computer security.

The main idea of this survey is to explain new trends in security, beginning with the primary concepts of security; cryptography and user authentication. Both explain briefly the main concepts of technology and how it works to maintain its features. The second part in every section explains what are the current trends in each investigation field, focusing on new technologies (quantum cryptography, visual cryptography, dynamic biometrical authentication) or the evolution of current research (Public key enhancements, elliptic curve cryptography, two-way authentication approaches) in order to find new ways to implement solutions and improve inherit vulnerabilities of every one of these technologies and their interactions with the users.

The security property known as availability is generally compromised after a Denial of Service (DoS) attack is launched. In the past, one or few devices were used to perform DoS attacks, but now, sets of infected computers or devices called botnets are used in a type of improved attack called Distributed Denial of Service (DDoS). Consequently, the attacks have evolved in a way that the attacker is practically undetectable. At the time, it is possible to find many techniques in which DoS attacks can be performed. They also use intermediary nodes to reflect or amplify attacks. In the same way, attacks have evolved, defenses and responses against attacks have also emerged. Technologies and concepts such as Software Defined Networking (SDN), granular computing, neural networks, machine learning, and feature selection have been used in related works.

Another essential aspect of security is intrusion detection. As soon as an intruder can be detected, the attack range can be limited or entirely avoided. This work presents a summary of some concepts of IDS, IDMEF, and honeypots, as well as recent studies to improve the performance in intrusion detection. Markov chains, multi–pattern string matching, signature matching algorithms, diffuse genetic algorithm are examples of methods based on intrusion detection. Additionally, firewalls, as the first line of defense in a network, are analyzed. The present work introduces firewalls concepts and a brief description of the main techniques to perform its action. In addition, it presents research advances focused on improve the effectiveness of traditional firewalls and challenges for cloud environments.

The rest of the paper is outlined as follows: Section 2 describes cryptography basis and new approaches. Then, Section 3 explains identity management and new manners of user authentication. Section 4 focuses on denial of service attacks, including DDoS, application-based, reflection and amplification. Section 5 introduces intrusion detection systems. Section 6 describes the firewall concept. Finally, Section 7 presents the conclusions and future work.

## 2. Cryptography

Cryptography is the most important principle to keep the privacy and information confidentiality, both analog and digital. The main idea behind this technology is, using some sort of information, keep illegible all messages sent and/or stored except for people who have access to this information, also known as a passcode. This simple idea has become the base of data security in the world. Last implementations are based on complex mathematical problems and our computational incapacity to solve them. The main uses for cryptography can be divided as folloes [1]:

- **Symmetric Encryption:** It is the classic cipher system, known since the roman empire. It consists of the use of a word or a combination of features to keep information secure. It is called symmetric because the cipher and decipher process depends only on a unique passcode that is mandatory to keep secret. All these systems, which are being based on human communication, are sensitive to statistic attacks (percentage of appearance of letters in a certain language). Since the Second World War (WWII) and Enigma cipher machine, symmetric encryption evolved in systems that convert messages in uniform strings of bits that keep secrecy beyond language, format, source or destination. This evolution led to secure algorithms like Data Encryption Standard (DES), Triple DES (3DES), the new Advanced Encryption Standard (AES) and their variants.
- **Public-Key Encryption:** This is the one true revolution in encryption systems, created in the late seventies. Instead of using a unique passcode to cipher and decipher, the new system uses two different although related keys. One of those are known (the public key) and the other one needed to keep in secret by the user (the private key). In this way a ciphered message with one of these keys is only deciphered by the other one. The first approach of public-key encryption was a simple but powerful mathematical calculation called the Diffie-Hellman algorithm, which depended on the communication of numbers which are easy to create but very hard to decompound in its original factors. Nowadays, the most common public key system in the world is RSA, but other techniques, like elliptic curve cryptography (ECC), are proven to be better and powerful systems than the classic ones.

  The public-private key systems allow not only cipher on-transit data (confidentiality), but also to check that on-transit message is in fact sent by the claiming author (the one who owns and keeps the private key). This feature is called digital signature, because of the ciphered text with the private key can be read for all the users who have the public component, which means that is easy to read. However, this deciphered secret text allows verifying the origin of the information.

In these cases, secrecy is not as important as the author verification. With all these features, it becomes mandatory to find a way to verify that person who owns a certain key pair is who claims to be. That need is the origin of Certificate Authorities (CA): big trusted entities who can provide and manage the authentication of all users. CAs' success lies in its capacity to verify the identity of all key pairs' owners and manage searches of keys and also revocation of all keys which could be compromised, lost or corrupted.

- **Hash Functions:** Cryptography provides the capacity to check that an on-transit message has not been modified since its creation. Among Digital Signature, Hash functions allow creating the same size digest of every message, in a way that the slightest modification in the message will produce two different digests. Together with the public-private key pair, the hash function provides confidentiality, authenticity, integrity and non-repudiation, also known as the four pillars of security.

### 2.1. Current Applications

Evolution of cryptography has focused on four big study groups which are current research fields.

- **Improvements on Public Key Systems:** there are two big problems with the improvement of Public Key Infrastructure (PKI): (1) eliminate unused or expired keys and (2) improve the use of all the system in order to prevent impersonation attacks or weak cipher algorithms. The first problem can be solved improving the access of CA's to systems where secure communications are mandatory, like SSL and HTTPS communications [2] and periodically cleans the CAs' caches and certificates. The second problem is a little more difficult to solve. A way to improve impersonation is to use a technique called Identity Based Cryptography [3]. This cryptography makes possible to implement a system that could be anonymous and improve the use of PKI strengthening algorithms like ElGamal.
  Other implementations of PKI systems can generate a set of functional private keys from a master key, which is also used for encryption in fields of randomized functions. The set of keys allows all the security features of multi-key systems based on identity [4]. The same idea could be extrapolated to generate two ciphered text from the same private key: a false ciphertext (used as a decoy and tag to open any message with consistent randomness, all part of a unique security system) and a real ciphertext which contain protected information. If a user doesn't have the proper key, he cannot distinguish the fake ciphertext from the real one [5]. Also, it is possible to study a new way to use PKI instead of the random creation of keys. A system called deterministic PKI [6] protect large amounts of information stored in cloud systems. These deterministic systems require a deterministic algorithm to create random public keys.
  Another implementation studies the keys in order to find out a encryption weakness as well as its relationship with certain plaintexts [7]. All these studies seek to reduce the vulnerabilities of PKI implementations for both, commercial and experimental fields.
- **Visual Cryptography:** It is a way to use cryptography without a computational system but the user's eyes. Users can solve a specific image from a set of incomplete images, where the overlapping can solve the system in the base of a coding table [8]. Figure 1(a) shows how a pixel can be used as a bit for the XOR operations which results in the final image, pixel-a-pixel. The uses of these operations are shown in Figure 1(b).

  This cryptographic system can be used to replace CAPTCHA or similar systems, to avoid the impersonation of humans by machines or artificial intelligence. However, both colored and black and white images have low-quality issues because the cipher is based on pixels. An improvement of this technique implies to use an XOR system where is not necessary a code table and works like steganography providing additional security in the transmission of secret shared information [9].

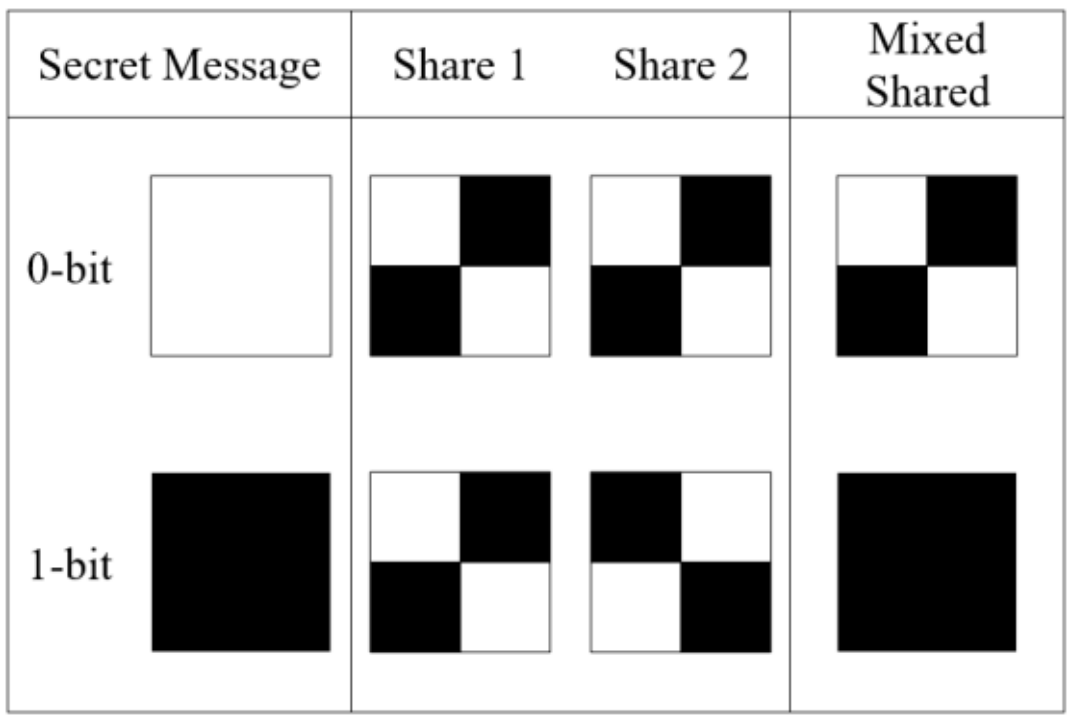

(a)

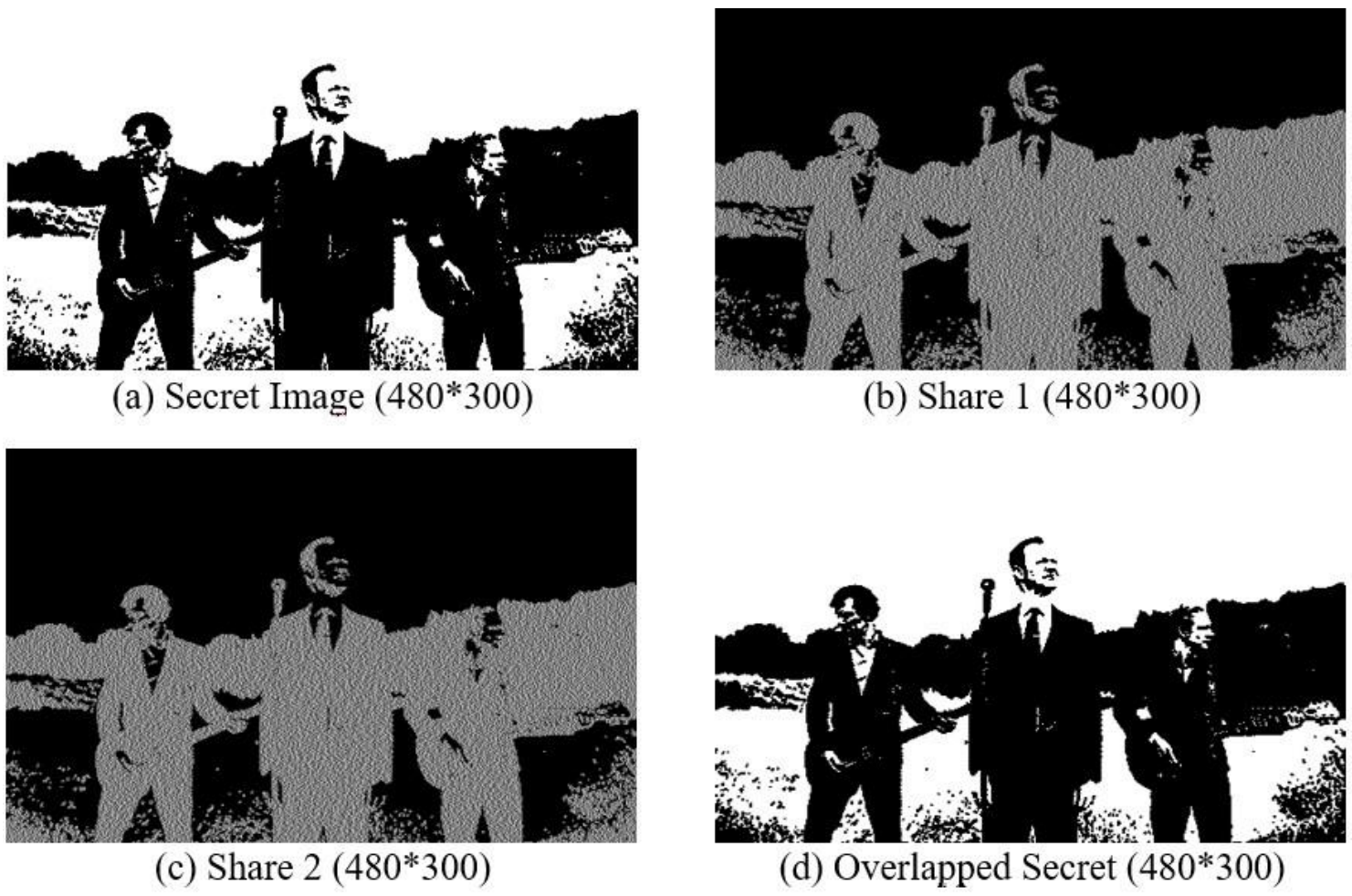

(b)

**Figure 1.** (a) VC Scheme to obtain a specified outcome for an image. (b) Monochromatic VC as a result of using the scheme in (a).

- **Elliptic Curve Cryptography:** One of the strongest mathematical systems to implement a PKI is ECC, because its key generation calculations are faster than traditional PKI's. For this reason, it is used mostly in mobile systems, like bitcoin wallets, but also can be used also in SSH servers, HTTPS certificates and validation and identification systems like LDAP [10]. Additionally, ECC is used in the improvement of the Internet of Things (IoT) [11], where using XOR protocols and a session key (generated from the curve), the communications improve the security between devices without all the resources consumption of RSA.
- **Quantum Cryptography:** Classic computation is still based on Shannon principles, the bit (a mathematical theory of communication and theory of secrecy systems, Bell Systems Technical Journal 27 and 28 respectively) and in the Moore Law. Stephen Weisner, in 1970, proposed the first ideas of using photons to transmit and storage information, based on the Heisenberg principle of uncertainly. It was the beginning of quantum computation and, in consequence of Quantum Cryptography (QC). The keys for these systems are created by the polarization and orientation of this polarization in photons, where orientation can represent classic bits and is part of a complex system known as quantum key distribution (QKD). In order to implement all the possibilities of QC, several protocols have been developed, like BB84, BB92, EPR, AK15, S13, among others [4]. All these algorithms could, theoretically, improve performance on cyphering and in avalanche effect than classic algorithms [12][13].

  These protocols could use to improve, the generation of public-private key pairs, improve one of the techniques known as coin tossing [14], without impersonation and forgery risks. Also could be used to improve Diffie-Hellman's like key generation for two users with BB84, as in Figure 2, where users don't choose prime numbers but photon polarizations and quantum channels. Another example of security improvement is the use of QC to improve IoT, where the security and performance of all devices won't be compromised by the algorithm [15]. Another technique is using a mathematical system of matrixes to generate keys in the function of a set of points where the matrix mix rows and columns, better known as Lattice-Based cryptography [16]. In this case, QC improves calculations in the Euclidian field for all points, with performance and energy efficiency.

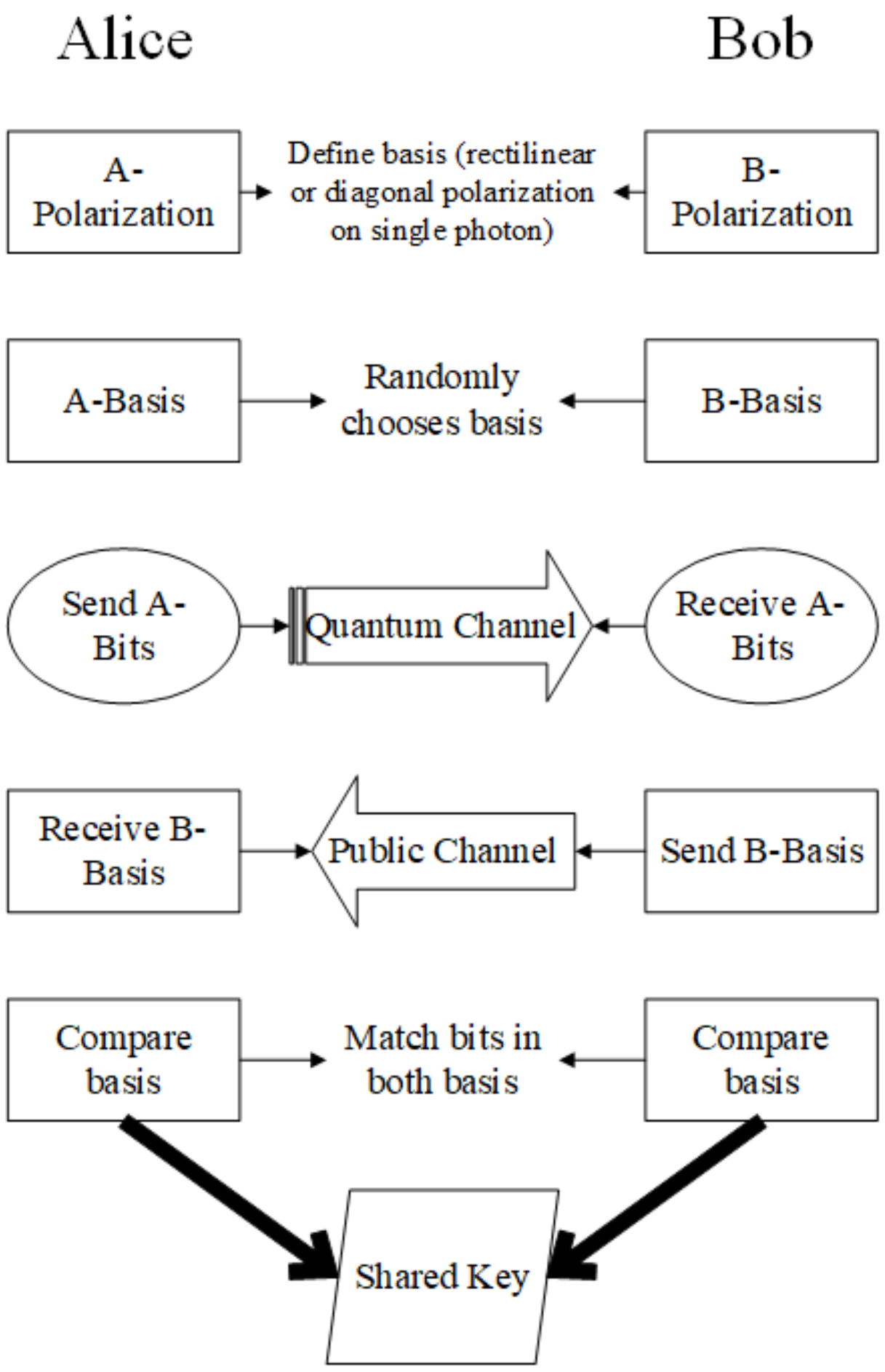

**Figure 2.** Steps of BB84 protocol's QKD.

## 3. User Authentication

One of the main uses of cryptography is the verification of the identity of users both in login schemes and communication via the Internet. Identity management is based on four main features, where users always must comply with at least one of them [1]:

- Something that user knows (passcode, as password-based authentication): uses a passcode to validate the existence of a user and if it is who claims to be. Besides all the flaws of this system, mostly for the users, are the main ways to determinate identity and authentication on the Internet.
- Something that the user possesses (physical keys or cards, as token based authentication): based on some sort of hardware that user possesses, which allow users to access physical places or validate its identity with specialized hardware.

- Something that user is (biometrical features, as biometric authentication): static biometrics allow a system to recognize a user in an unequivocal way, using biometric values that are unique. These features are more accurate than passcodes but are also expensive.
- Something that user does (patrons of biometric data, as dynamic biometric data): an evolution of static biometric authentication, is based in how the human body interacts with its environment, where certain patrons are as unique as static biometric data, like typing speed, handwriting, voice pattern or brainwaves.

The interaction of one of more of these features improves the authentication of users in all systems, especially because biometric data cannot be forged.

### 3.1. Current Applications

All evolution of user authentication is based on improving at least one of the previous features in a way that cannot be forged or manipulated in transit or authenticated systems. There are two big research fields in authentication, which are:

- **Two-way password related schemes:** Passwords have been evolved to apply in all systems that need user authentication [17], especially in mobile devices. Using this idea, several systems have been designed to improve mobile security, which have a problem of weak and short passwords. In [18], a virtual dynamic keyboard was generated. It varies the positions of every character and/or number in every interaction. Another system uses graphic patrons that user must recognize in a preset code database like a dynamic pin [19]. The solution in [20] uses a One-Time Pad scheme for authentication through a mobile device, which is connected with a user management server. This scheme uses all the cryptographic communication features to establish a secure channel avoiding the need to enter usernames or passwords by the user. Another use of these technologies is to generate a pseudo password which is a blend between a formal user password and a bitmap image features (an image chooses by the user) in order to create a specific derivation password to access to certain systems [21]. This scheme implements a two-way authentication system where passwords and their relationships are the access key and the authentication scheme at the same time. However, password-only systems are still the simplest schemes to authenticate users, and in this scheme, the application of the correct system and the correct password complexity are the strength of all of it, besides its simplicity [22].
- **Dynamic biometrical related schemes:** All schemes use new ways to read dynamic biometric features, or new dynamic biometric features, because all the signals that belong to the human body are unique and could be used to authenticate the user. The importance of all these features lies in, besides other systems, users cannot forget or lost his own biometrics. One of the most ambitious systems uses brain waves to improve authentication, using bluetooth sensors to find a patron in brain waves that identifies a unique user [23]. This system is still in development but could be an important way to determinate identity. Other systems, designed for mobile devices, intent to determinate patrons of use of the mobile's owner (unlocking, typing, browsing habits, passcodes or PINs among others) [24]. The systems can use a combination of patrons or just one of them, while implements a dynamic biometric system with password patrons [25] or access phone patrons [26]. Both experimental systems are the proof of biometrical patrons while mobile devices are used. Another scheme implements multimodal dynamic biometrics, in order to blend two or more biometric patrons [27]. Other systems intend to use big data to determinate all the user's biometric patrons. This system can create a patron that identifies a user, with databases and connections that also works with biometric schemes. The idea is to implement a new framework to distribute and standardize dynamic biometric schemes [28,29].

## 4. Denial of Service

A DoS tries to flood a resource by sending useless information until it can't respond to legitimate users requests. It can attack network bandwidth, systems, and application resources. A DoS can be performed using techniques such as flooding ping and SYN spoofing. Flooding ping is the most basic DoS attack, it is done by sending a huge number of ping commands. The ping of death sends ping command with the largest packet size. A SYN spoofing is a common DoS attack, which is based on the handshaking process. A SYN flag with a spoofed address is sent to the victim, and it sends a SYN-ACK flag to the spoofed address many times until the connection request fails. In general, a SYN spoofing attack is the combination of IP spoofing and SYN flooding[30]. There is an improved DoS called Distributed Denial of Service (DDoS). It commonly uses a botnet to perform the attack instead of a simple host or few ones. A botnet is a big set of infected computers or devices (zombies) that can be controlled remotely by an attacker. According to the reports of Verisign[31,32], the number of DDoS attacks has suffered a considerable increasing compared to the same previous quarterly report. Meanwhile, SecureList[33] established that more than half of the total number of attacks are SYN flood type. A brief summary of these information is shown in Figure 3. The most important types of DoS attacks are described briefly below.

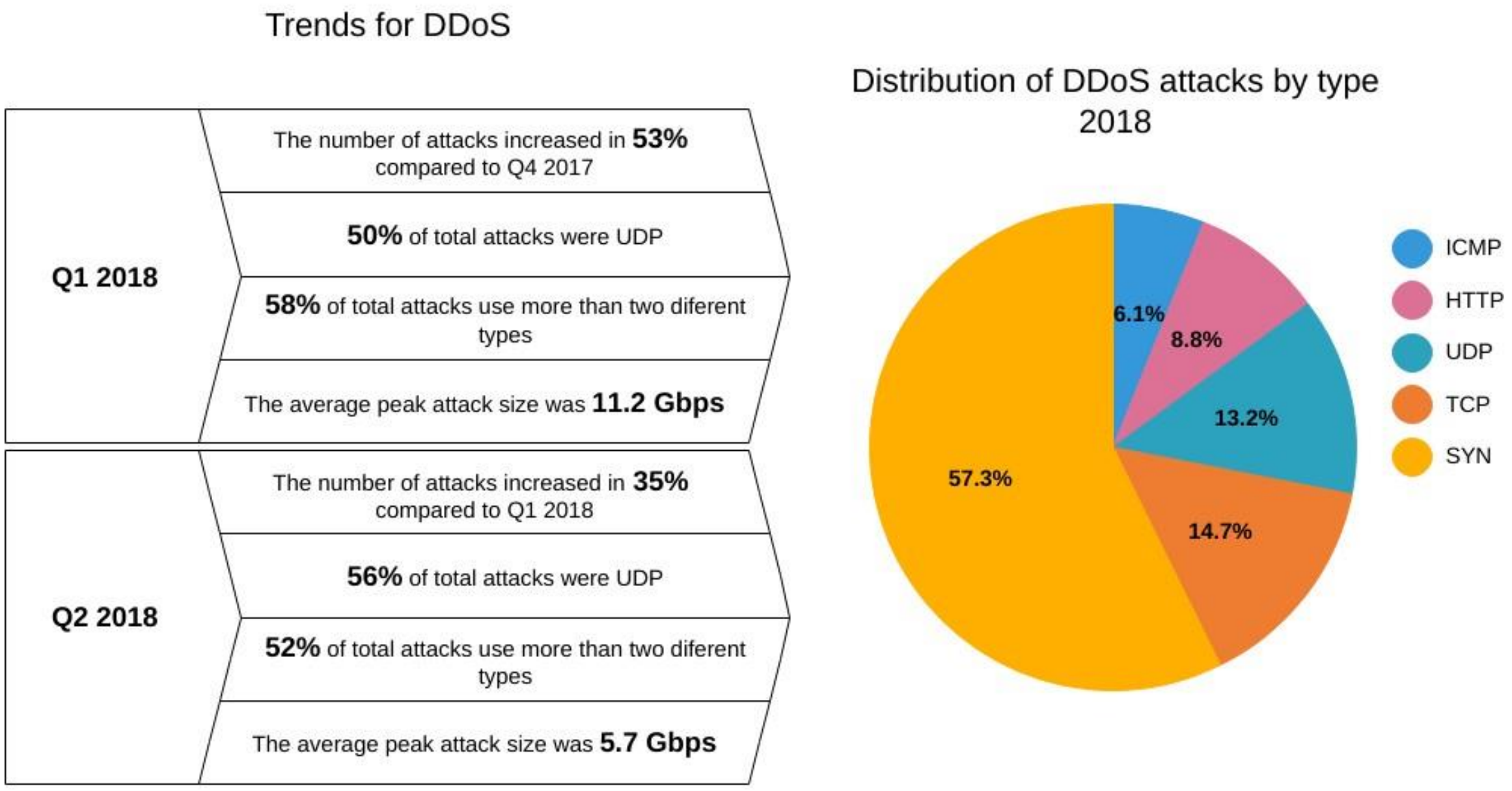

**Figure 3.** Trends and distribution of DDoS attacks in 2018[31–33]

- **Flooding Attack:** A flooding attack is also called a direct attack because there is no intermediary node between the attacker and the victim[34]. The most common types are ICMP, UDP and SYN flood. In the ICMP flood, a great quantity of ICMP packets of different size is sent to the victim in order to receive response packets. During the performing of a ping command, an ICMP echo request is sent by the attacker and an ICMP echo reply is sent back by the victim. In the UDP flood, the operation is similar to ICMP flood but, UDP packets are used instead of. Finally, in the SYN flood, TCP packets with SYN flag enabled and spoofed source address are sent. For its part, the victim sends packets with SYN/ACK flags and wait for the packet with ACK flag, but it never happens because of the packet contains no real address.
- **DDoS:** A DDoS is composed of 4 components: (i) the target or 'the victim'; (ii) agent programs installed without authorization on the victim and hosts called 'attack daemon agents'; (iii) the program that will conduct the attack or 'the control master program'; and (iv) 'the attacker'. The attacker communicates with the control master program to coordinate and launch the attack through the following techniques: Trinoo, TFN, Stacheldraht, Shaft, and TFN2K [35]. For instance, TFN2K uses TCP, UDP or ICMP to communicate the control master program with attack daemons agents. Finally, the attack daemons agents can launch the attack using SMURF, SYN, UDP, and ICMP flood attack. At the time, DDoS attacks have suffered improvements in performance, operation, and motivation. For instance, the Internet currently offers services to perform DDoS attacks for money as a business, they are commonly called 'Booters' or 'Network Stressers' [36]. Mirai is a botnet, which has been used to launch the most powerful DDoS attacks. Mirai botnet is composed of 4 components: the bot, which infects IoT devices and also attacks the target after a command is sent by its botmaster; the command and control (C&C) server provides the interface to coordinate the attacks; the loader delivers executables to new victims, and the report server provides information about the devices in the botnet. The components and operation key steps of a Mirai botnet are shown in Figure 4.

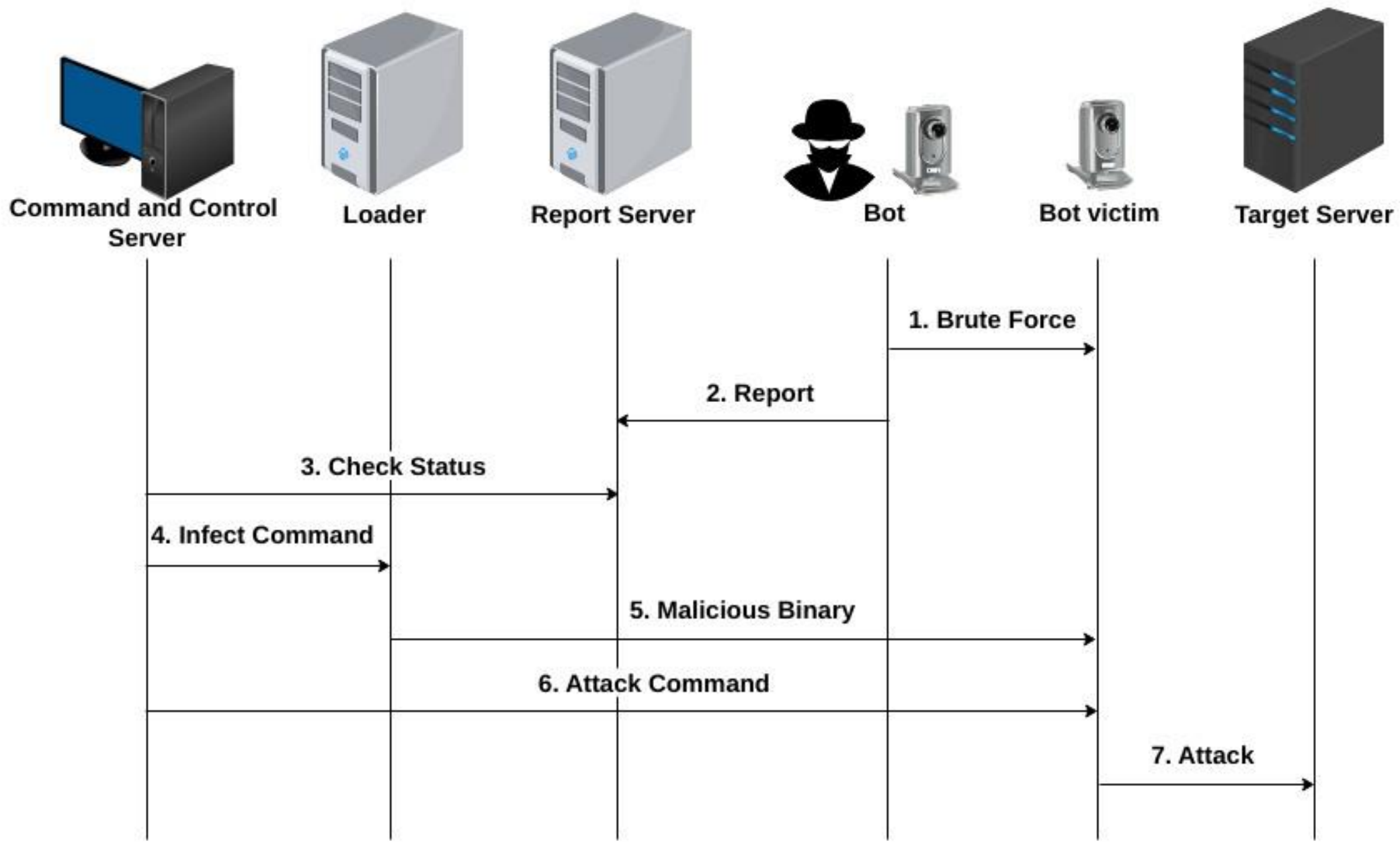

**Figure 4.** Mirai operation and communication

- **Application-Based Bandwidth Attack:** An Application-Based Bandwidth attack tries to force a target to expend a huge amount of resources using the nature of the protocols. It is made by exploiting the vulnerabilities of them. Two of the most popular applications on the Internet are the World Wide Web (WWW) and Voice over IP (VoIP). In this way, the HTTP flood and SIP flood attacks are designed and used to perform attacks. For their part, firewalls have been forced to keep open ports and permit HTTP/HTTPS traffic. In general, an HTTP flood attack is performed against a web server using HTTP requests. To conduct an HTTP request to a web server, a TCP connection has to be established, therefore, a genuine IP address is also required. The attacker instructs bots the way the attack will be performed. Its aim is to make sure the target incurs in a high resource consumption avoiding the attack traffic is cataloged as malicious. In VoIP, a protocol known as Session Initial Protocol (SIP) is used to establish a call. The idea of a SIP attack is to flood SIP proxies with many invitations. Like HTTP flood, a botnet can be used to send invitations and avoid anti-spoofing mechanisms. In SIP flood attacks, SIP proxies as much as call receivers can suffer the attack.
- **Reflector and Amplifier Attacks:** They are also called indirect attacks because there are intermediary nodes between the attacker and the victim. The intermediary nodes (reflectors), can be routers or servers. First, the attacker sets the victim's address as the source address of its packets. The packets with spoofed source address are sent to the reflectors. The reflectors naively respond to those packets sending response packets to the victim. Depending on the number of reflectors, this attack could flood the reflector-victim link. Smurf is the name given to a classic reflector attack[34]. On it, spoofed packets with the victim's address are sent to the broadcast address, the hosts that receive those packets respond to the victim, this back and forth, finally, it generates enough traffic to flood the link of the victim. Reflector attacks can use any protocol that allows generating automatically responses to the messages. Unlike reflector attacks, amplifier attacks use a botnet instead of a single host. The attacker sends a command to every bot or zombie. For its part, they launch an amplifier attack against the victim using reflectors. The architecture of a reflector and amplification attacks is shown in Figure 5.

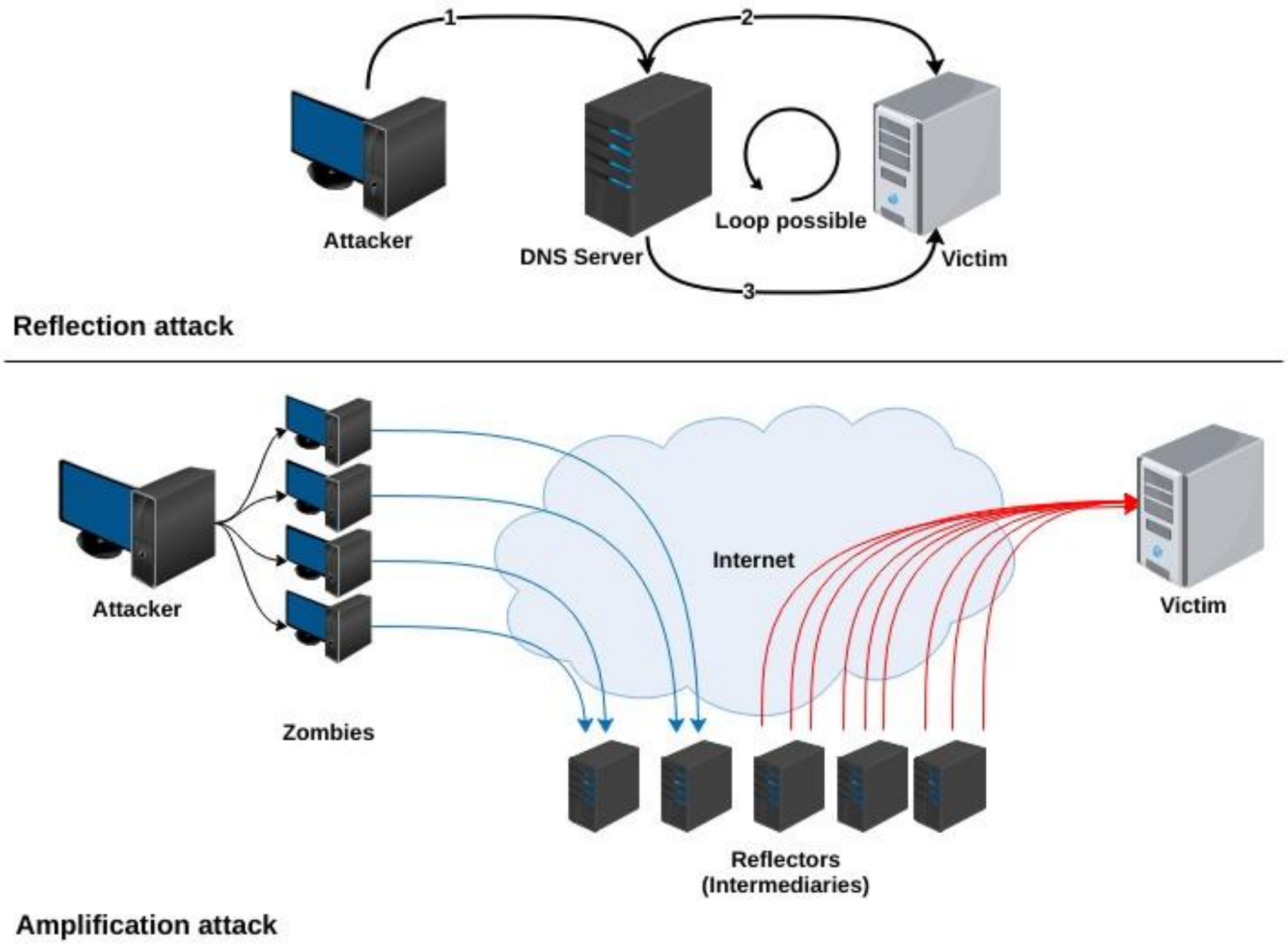

**Figure 5.** Reflection and Amplification Attacks Architectures

### *4.1. Related Works and Trends*

Due to the popularity and spreading of the Internet of Things (IoT) and cloud computing, DDoS attacks are already present on them. Kolias et al.[37] advertised the emerge of IoT DDoS attacks few years ago and they proposed a review of the botnet called Mirai and its variants. On the other hand, Ficco et al.[38] proposed a new generation of DoS attacks focused on IT cloud computing, in which the main purpose is to increase the normal operation of specific equipment. Several variants of Mirai bot have been created after the release of its source code. In spite of this fact, IoT vendors have shown little interest in improving the security of their devices. For this reason, IoT devices are been still infecting even using the original version of Mirai. As it is expected, many variants of Mirai bot appeared, each one providing new features and improvements. BrickerBot [39] attempts to generate a Permanent Denial of Service(PDoS) attack using different techniques such as modifying IoT device firmware. The idea is to make certain behavior as permanent (low level of security, low in maintenance, the capability to launch an attack, and limited user interfaces).

The vertiginous growth of cloud computing has promoted that enterprises migrate their infrastructure to the cloud. DDoS attacks have improved because of this trend. In [38], the authors proposed a type of attack named energy-oriented Distributed Denial of Service (eDDoS). Its main purpose is to increase the normal operation of the IT and Heating, Ventilation, and Air Conditioning (HVAC) equipment instead of blocking them. In other words, eDDoS tries to increase the power consumption of infrastructure in order to reduce the components lifetime. Modern IT infrastructures have energy efficient hardware components, so, they can operate using different power levels. It means, they can switch dynamically between lower and higher power levels depending on their workload. eDDoS takes advantage of this feature to make certain equipment always operates at higher power levels.

The HTTP/2 standard has been conceived to support high-speed communications. It introduces some features such as binary framing and application layer flow control. HTTP/1.1 traffic patterns are used by current DoS detection techniques against web servers. However, the use of HTTP/2 traffic patterns can be practically null. In this context, [40] two stealthy DoS attack models were proposed (SA-1 and SA-2). The stealthy models attempt to hide traffic attack within legitimate traffic. Both use an HTTP/2 feature called flow control that advertises congestion issues. This feature moderates the streams generated between a client and a server. There is an HTTP/2 message that implements flow control known as window_update. It is used to notify to devices the data size that sender can transmit in one frame. According to [40], there are two studies related to DoS attack modeling against HTTP/2 services, both use the window_update frame. Similar to previous works, the present models also use the window_update frame. On the one hand, SA-1 includes tro groups of bots, the mime group and the offending group. The first one is used to mimic legitimate traffic and the other one to generate attack traffic. The mime and offending group use some parameters in order to generate legitimate and attack traffic. One list of five parameters used by both is shown in Table 1. The simulation results of SA-1 showed a total consumption of CPU resources with two bots. On the other hand, SA-2 uses four bots. Bots 1 and 2 are the mime group and bots 3 and 4 are the offending group. In this model, Bots 3 and 4 don't send windows_update traffic to the victim. The parameters of the SA-2 model are shown in Table 2.

**Table 1.** SA-1 Model Parameters

| Parameter | Bot 1 | Bot 2 |
|---|---|---|
| Number of threads | 1 | 1 |
| Number of window_update | 131 K | 131 K |
| Stealthy factor | 50 | 500 |
| Delay between connections | 11 ms | 11 ms |

**Table 2.** SA-2 Model Parameters

| Parameter | Bot 1 | Bot 2 | Bot 3 | Bot 4 |
|---|---|---|---|---|
| Number of threads | 1 | 1 | 2 | 40 |
| Number of window_update | 131 K | 131 K | 0 | 0 |
| Stealthy factor | 50 | 500 | N/A | N/A |
| Delay between connections | 1 s | 1 s | 0.001 ms | 5 s |

With the arrival of network stressers or booters, such as attackers as defenders have been forced each other to implement security infrastructure, attack detection and mitigation schemas, and attack models. Krupp et al.[41] proposed a strategy to outlaw the use of booters by attributing amplification attacks to booters services. The term used by some booters to proclaim their services as legitimate is stress-testing services. This approach proposes a classifier based on k-Nearest Neighbor (k-NN) classification algorithm. Using attack datasets provided by honeypot operators or victims, three key features were identified (i) honeypot sets, (ii) victim ports entropy, and (iii) time to live (TTL) values. Honeypots are used by booters as reflectors. (i) According to observations, honeypots seem to be reused against different victims. (ii) Despite the attacker can assign different ports to the victim, the attacks use one or very few victim ports. It could be because of booters ask their clients for one victim port. (iii) One or very few TTL values are shown by honeypot, therefore, it could be concluded that most attacks come from one source. These features are used for training and validating the classifier. Considering that datasets could miss data for some boosters, every attack can't be attributed to one booster. In view of this inconvenience, the strategy included a threshold. It refers to the radius covered by the neighborhood. In the case of a certain item is not eligible, it would be classified as unknown. Finally, with a precision of over 99%, DNS and NTP attacks can be attributed to booters using the classifier. Also, from a set of attacks observed by honeypot operators, 25.53% of DNS attacks could be attributed to 7 booters and 13.34% of NTP attacks to 15 booters.

Despite flooding ping and SYN spoofing are the most basic attacks, there is a current approach for detection and mitigation of these types of attacks. Kavinsankar et al.[42] proposed an Efficient SYN Spoofing Detection and Mitigation (ESDMS) Schema. The ESDMS is the combination of the Efficient SYN Spoofing Detection Schema (ESDS) and the Efficient SYN Spoofing Mitigation Schema (ESMS). ESDS has an attack detector component composed of a TCP probe generator and a filter. The architecture of ESDMS is shown in Figure 6. The detector component receives a packet with the SYN flag and Probe Generator sends back client some probe requests. When the source address is real, the probe requests are answered immediately. ESDMS provides a confidence table which includes IP addresses and trust values. Also, threshold values are compared to trust values to determine spoofed addresses. TCP probing method is used to calculate trust values, and statistical testing to threshold values. On the one hand, packets with addresses identified as trusted ones are accepted and these are included on a filter, and on the other hand, packets with the spoofed address are rejected. The filter contains a list of updated trusted addresses. It also is periodically flushed. ESMS checks the presence of source address on the filter list and its availability using TCP probing to permit the connection establishment. Otherwise, when the availability of a source address fails, a SYN spoofing alarm is generated. ESDMS shows better performance than other solutions such as SYN cookies, HOP count and IP puzzle.

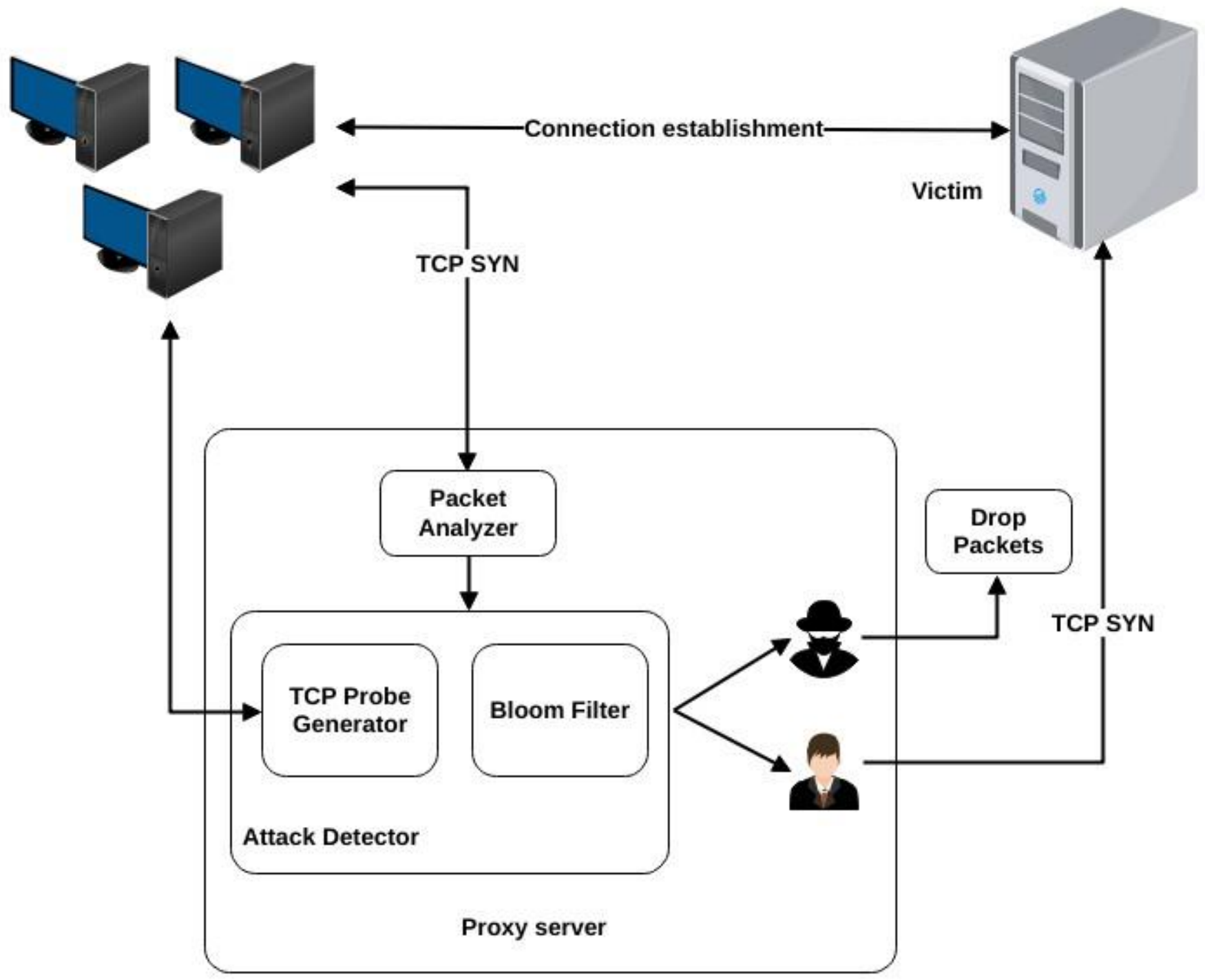

**Figure 6.** ESDMS architecture

In view of pervasiveness of technologies such as SDN, some approaches to mitigate the effects of DDoS attacks have been proposed. Mohammadi et al.[43] proposed a countermeasure based on SDN to mitigate SYN flooding attacks called SLICOTS. Meanwhile, Prakash et al.[44] proposed an intelligent SDN based on machine learning algorithms for preventing DDoS attacks. SLICOTS detects and mitigates SYN flood attacks in an SDN architecture. It was conceived like a module for security and implemented in the control plane. The SDN controller platform used by it is OpenDaylight[45]. Its architecture is shown in Figure 7 and only works in reactive mode by creating forwarding rules by the time the attack happens. One path for forward and one for backward between the client and server is established. It is applied to every TCP session. SLICOTS listens in progress process of handshaking and installs on OF switches short-term forwarding rules. If the handshaking process is completed, the short-term rule is replaced by a permanent one. Otherwise, and considering a threshold for half-open connections, a rule for blocking the client is created on the switch. Thus, the switch will block malicious hosts and avoid the controller to get saturated. The key success of SLICOTS lies on two parameters the number of hosts in the network and the number of illegitimate requests for a specific client.

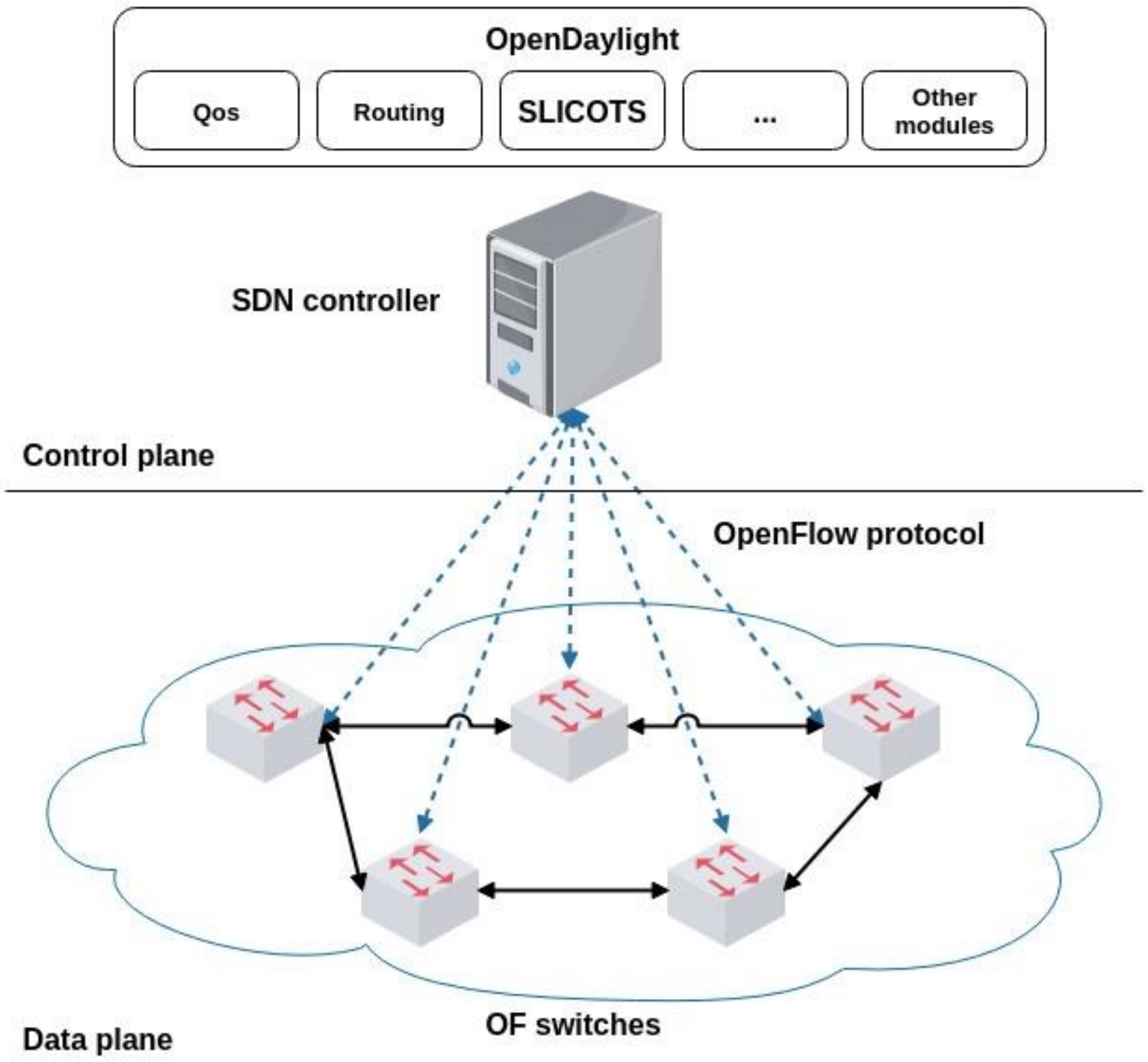

**Figure 7.** SLICOTS architecture

An Intrusion Detection System (IDS) is one of the most common components to detect different types of attacks. An IDS has two types of operation (i) anomaly intrusion detection and (ii) misuse intrusion detection. On one hand, anomaly intrusion detection uses normal usage patterns. Those patterns are built using measures of system features. On the other hand, misuse intrusion detection uses malicious behavior patterns (signatures). IDS design is based on neural networks, support vector machines (SVM), and fuzzy logic. Feature selection is a key concept in IDS because it finds an efficient subset of features to make sure the accurate prediction. The dataset NSL KDD used on this approach consists of 41 features of network traffic. It is applied for anomaly detection; thus, it covers DoS attacks. In [46], the authors propose a way to select DoS attacks features using entropy and granular computing. It is based on Shannon entropy to calculate the weight for features, and, granular computing to select potential features/attributes. Table 3 shows a summary of each selected features with values of probability, entropy, and weight. Meanwhile, a brief summary of potential features is shown in Table 4. In comparison to some proposed methods to detect intrusion using NSL KDD dataset, the current method identifies DoS attacks potential features instead of only features. It will help IDS to focus on these potential features to detect the attack, being an option to implement this approach on SDN [46].

## 5. Intrusion Detection

In computer security intrusion is a set of actions that can compromise the integrity, confidentiality or availability of a resource. Nowadays, there are several types of intruders: hackers, crackers, sniffers, spammers, illegal suppliers, scriptkiddies and phreakers. They may cause an individual or organized attacks with different skill levels, tools and resources [47]. If the objectives and attack fields are identified, the vulnerabilities can be determined.

Internal intruders may change the information intentionally or involuntarily in a host, network, or service, which may cause theft, modify or removing information as well as computer sabotage in an organization. External intruders send files (programs, computer applications) to a network or host with the purpose of adding or modifying data in legitimate user, access controls or server without authorization. IDS allows detecting unwanted or anomalous actions inside or outside the computer system, acting on hosts (HIDS), networks (NIDS, DIDS). Also, these analyze behavior users, knowledge acquired, and generate reports.

Currently, IDS provides greater security, autonomy, compatibility, failover, without overloading the host or network. They are compatible, upgradeable, as well as fulfill prevention function with virtual sensors (sniffers) and defense before the attack [48]. However, IDS can generate false positives when the system mistakenly indicates an intrusion and can block a part of it. Otherwise, the system can generate a false negative and the system does not detect an intrusion.

HIDS analyzes behavior in files, resources, audit trails, system logs and application logs within the network [48]. HIDS resides on the same host without affecting network performance and includes a knowledge base of attack types and variations that allow them to be updated.NIDS detects anomalies in the network, through real-time analysis of incoming and outgoing traffic, comparing suspicious patterns and generating alerts, it does not affect the network performance, but it requires own hardware for processing (ex. SNORT). DIDS analyzes the traffic in a network and hosts and it also allows to manage the system alerts. It simplifies intrusion detection, analyzing the network traffic before it happens, however its disadvantage is the high computational cost. The central administrator associate system events and matches the signatures of intrusion.

Intrusion detection analysis [49] can be classified in two types principally, anomaly detection and signatures, as indicated in the following Table 3:

**Table 3.** Intrusion Detection, Analysis Approaches.

| | Behavior | New Attacks Detection Detection | Methods | Ex. |
|---|---|---|---|---|
| Anomaly detection | Search differences | Yes | Static tools<br>Knowing (set of rules)<br>Algorithms (Machine Learning, training data) | – |
| Signatures | Search likeness | No | Signature comparison<br>Heuristic rules (match methods) | SNORT |

Intrusion Detection Exchange Format (IDMEF) is a protocol based on format definitions and procedures for information exchange of alerts between devices or systems. Messages exchange uses data models in XML, JSON and SMI method for their equivalent representations [50]. The Internet Engineering Task Force (IETF) continues setting the process of intrusion detection messages [51].

A honeypot is a security tool in a network used to trick intruders (trap), detecting their attacks and then recording data results from system vulnerabilities, analyzing activities and attack effects. A honeypot can give alerts when an attack occurs (Low interaction) and explores files (High interaction) searched by the attacker (system logs, identifies the IP address of the attacker). In addition, it can delay the reaction time of the attacker.

### *5.1. Related Works and Trends*

Recent studies show that improvements are being made in the performance of IDS [52], through signature matching algorithms to identify internal attacks, and a diffuse genetic algorithm method for external attacks detection. These are some methods to mitigate attacks on a system and reduce potential threats.

Other research [48] authors analyzed and mentioned NIDS for improving their performance (accuracy, detection rate) and decrease the false alarm rate. The IDS should be combined with a recurring feature selection method (RFA) to find interdependent functions that reflect several possible scenarios. In addition, the bigram technique should be used to codify these features and prepare them for using in machine learning and codify the "payload features" to avoid over adjustments due to data scarcity, in case of "zero-day exploits". For tests, the "ISCX 2012 data set" was used, which includes real traces analyzed to create profiles for agents that generate real traffic for HTTP, SMTP, SSH, IMAP, POP3, and FTP.

Currently, several investigations have proposed improvements for the honeypot implementation and their performance [53] in order to create defense scenarios with Firewalls, IDS and honeypots.In this context, performance evaluation is done through probabilistic analysis with the Markov chains. In another study, the model Hidden Markov Mode HMM [54] was used to determine advanced attack state sequences and their adaptation in an IDS. The performance of this doubly stochastic process in a testing environment and in a local network has been better than the decision-based approach and the neural network. False-positive alerts are substantially reduced.

"Middlebox applications" (IDS, firewalls, traffic classification, network censorship) use pattern matching to analyze the traffic flow in the network. When a deep packet analysis is performed, the performance of the system is affected by the computational cost required by these processes. A study on "Multi-pattern string matching" [55] has proposed an algorithm called DFC, with better performance than the "classic Aho-Corasick AC algorithm" used by tools such as Snort, Suricata and web firewalls (ModSecurity, WAF). The authors verified through laboratory tests and real applications that the DFC algorithm has reduced the use of memory resources and cache failures. It classifies and handles multiple strings of patterns by size and applies multiple progressive filters. It has been applied in IDS, ISP and firewall environments improving the performance of traffic classification and effectiveness of the antivirus, verifying that pattern matching modules can be replaced by DFC.

Other authors [53] proposed a method to evaluate the effectiveness of honeypots, using different defense scenarios through Stochastic Petri Net SPN, which is analyzed with Markov Chain (MC) processes, due to their similarity. The method was structured by system states, system transitions and the connection between them. With this proposal has been possible to extend the consumption time of the "Honeypot" during the defense and protection of computer systems; therefore, the performance was improved and evaluated.

## 6. Firewall

Firewalls are used to examine and decide if each packet must be allowed or rejected, its function is traffic regulation by trust level: higher (internal network), lower (internet) or intermediate (DMZ). A firewall represents the first line of defense in a network, stops attacks, optimizes network availability and prevents the reprocessing of illegitimate requests. The implementation of a Firewall can affect the internal network performance when applying different security policies in an organization [56]. The firewall must be configured through an ordered set of rules (destination IP, source and destination ports, protocols), each one is associated with an action (accept, register, reject). The goal of these rules and actions is to allow legitimate traffic and block unwanted traffic. The firewall itself has to be immune to penetration. Only authorized traffic is allowed, based on the local security policy.

Access policies must contain types of content and authorized traffic. Previously, an information security risk assessment must be carried out considering the policies of the organization. Filtering process includes: IP address, application protocol, user identity, and network activity. Firewall types are classified by its functions in a system; these are packet filtering, stateful inspection, application proxy and circuit-level. Firewall topologies can be: host-resident firewall, screening router, single bastion inline, single bastion T, double bastion inline, double bastion T, and distributed firewall configuration.

Intrusion prevention system (IPS) is an IDS that acts as a firewall. There are two types: HIPS and NIPS. Host based IPS (HIPS) has a HIDS, detects and analyzes it with isolation (sandbox) and can take actions. Network based IPS (NIPS) has a NIDS, if it finds a threat it can block traffic [57]. SNORT INLINE: allows blocking traffic, the evolution is called IDS + IPS + NSM (Suricata). Another security solution is Unified Threat Management (UTM) contains multiple security features integrated into one box that act at the same time to protect the system [58]. However, they require updates, the sizing of the equipment is proportional to the network and the services provided by the company.

### *6.1. Trends*

Nowadays, improving firewalls is an option adapting to cloud and SDN networks with new architectures. Recent studies analyze vulnerabilities within traditional firewalls and cloud environments. A study of SDN based Firewalls [59] proposes an extension of Science DMZ design, with NFShunt based on Linux Netfilter combined with OpenFlow switching. NFShunt is included as part of the Firewall's rule set (IP-Tables) with an additional control plane. A high-end Cisco firewall, free / open source software, and OpenFlow software were used for the study. The results were evidenced in the improvement of the Firewall performance in a test with users that demand a lot of data (near 10 Gbps). With this hybrid firewall, the inverse relationship between firewall security and network performance is reduced.

Cloud Firewalls (Virtual Firewall) demand additional characteristics to traditional networks. A recent study [60] proposed the requirements for a hybrid and distributed IDS as well as security levels for virtualization layer. The challenge is to minimize the communication and computational overload used between several cloud hosts. VF allows to apply security policies when they migrate from one host to another one.

IPS controls network traffic and blocks intrusions in real time. In [61], the authors proposed an IPS based on the cumulative sum (CUSUM) algorithm called as CSIPS used to prevent DoS and DDoS attacks. Packets are duplicated and sent to the IDS that detects malicious packets through CUSUM, which identifies the source IP address in the HASH table. The study summarizes the firewalls and their limitations, as is shown in Figure 8.

| Firewall type | Summary | Limitations |
|---|---|---|
| Packet-Filtering | Predefined set of rules | Not complex rules |
| Stateful | Keep track | Complex than first type |
| Proxy | Based on rules | Slower than first type |
| Web Application | SQL injection attacks | Outgoing network |
| Virtual | Virtualized environment | VM configuration |

**Figure 8.** Types of firewalls and their limitations [60]

Intrusion detection and prevention systems (IDPS) represent the second defense against intrusions. IDPS blocks malicious traffic reports at the administrator and removes corrupt packages. In [61] the relation between traditional IDS / IDPS and cloud is presented (Figure 9).

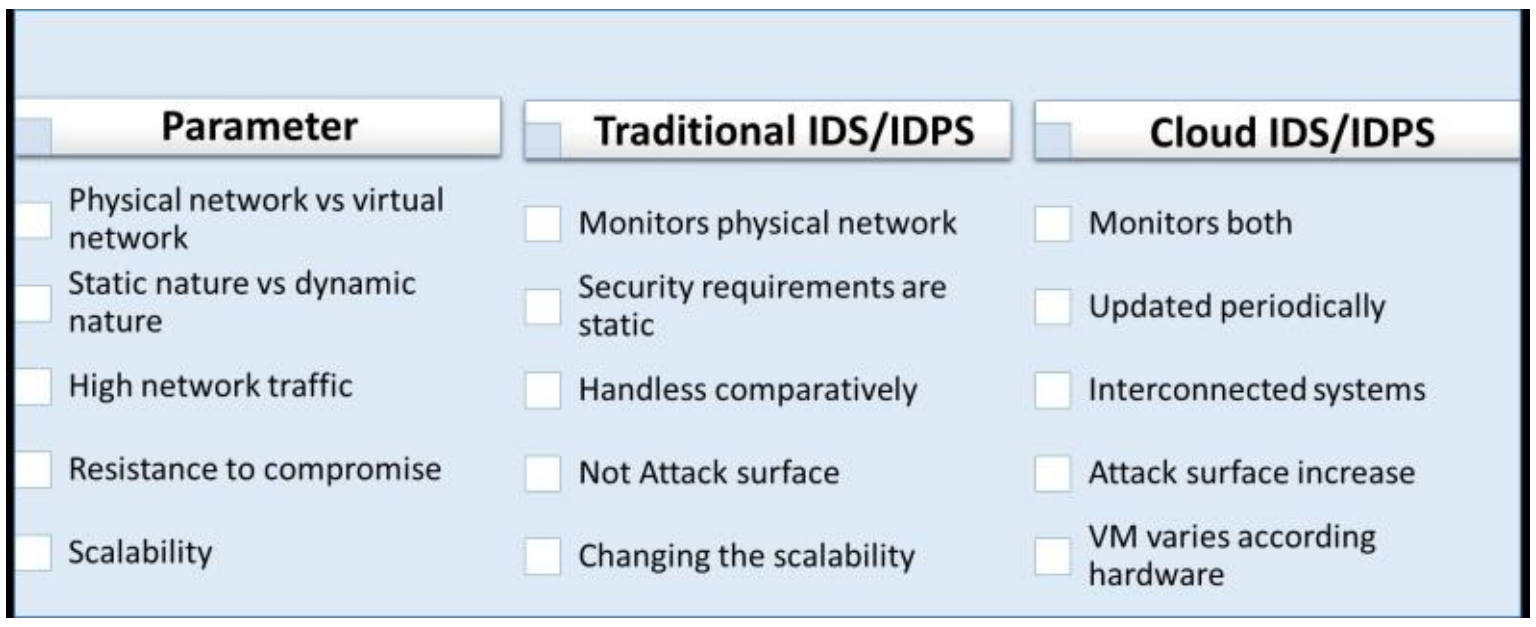

| Parameter | Traditional IDS/IDPS | Cloud IDS/IDPS |
| --- | --- | --- |
| Physical network vs virtual network | Monitors physical network | Monitors both |
| Static nature vs dynamic nature | Security requirements are static | Updated periodically |
| High network traffic | Handless comparatively | Interconnected systems |
| Resistance to compromise | Not Attack surface | Attack surface increase |
| Scalability | Changing the scalability | VM varies according hardware |

**Figure 9.** Comparison between traditional and cloud IDS-IDPS [60]

## 7. Conclusions

Evolution of cryptography and identity management shows that technology has evolved in a manner that cipher/decipher calculations need to be very small and very powerful. All applications, for secure information and identities, need to be functional for small devices specially in IoT terminals or personal mobile devices. Most of all current technology intends to improve systems and techniques where security doesn't depend on user-defined passwords, but keys and passcodes generated by a third party, like a CA or defined biometric patrons. This survey shows how mobile systems and biometrical patrons are the new ways to implement security and privacy for all users, in all stages of current technology.

The countermeasures to defend and respond to DoS attacks have evolved in function of the appearance of new attacks and their consequences. The idea is that the countermeasures go a step forward to them. Despite, the response to attacks is fundamental in information security, this field has not received the attention that it deserves. At the time, research efforts have been focused on design and implement mechanisms for prevention and detection but not for responding attacks. The present work presented some approaches about improvements of attacks and countermeasures, such as SDN, granular computing, neural networks, machine learning, feature selection, fuzzy logic among others. Similarly, it is also important to highlight the growing IoT market and the discovery of serious vulnerabilities on IoT devices.

Finally, it is worth mentioning that current researches on intrusion detection and firewalls show a greater tendency for correlating and integrating common methods and solutions. It has been demonstrated that machine learning oriented to rules can help to reduce false positive rates and develop advanced systems based on the behavior of new attacks, with the goal of achieving more accurate results. Another trend is the use of stochastic processes (Markov chain) to analyze new attacks.

**Author Contributions:** "conceptualization, writing–original draft, writing–review and editing,P.M., D.M., S.A., L.B., and A.V.", please turn to the CRediT taxonomy for the term explanation.

**Funding:** This research received no external funding

**Conflicts of Interest:** The authors declare no conflict of interest

## Abbreviations

The following abbreviations are used in this manuscript:

| | |
|---|---|
| 3DES | Triple Data Encryption Stardard |
| AES | Advanced Encryption Standard |
| CA | Certificate Authority |
| CAPTCHA | Completely Automated Public Turing test to tell Computers and Humans Apart |
| CUSUM | Cumulative Summation algorithm |
| DDoS | Distributed Denial of Service |
| DES | Data Encryption Standard |
| DIDS | Distributed Intrusion Detection System |
| DMZ | Demilitarized Zone |
| DoS | Denial of Service |
| ECC | Elliptic Curve Cryptography |
| eDoS | Energy Oriented Distributed Denial of Service |
| ESDMS | Efficient SYN Spoofing Detection and Mitigation |
| HIDS | Host Intrusion Detection System |
| HMM | Hidden Markov Model |
| HTTPS | Secure Hyper Text Transfer Protocol |
| HVAC | Heating, Ventilation and Air Conditioning |
| IDMEF | Intrusion Detection Message Exchange Format |
| IDPS | Intrusion Detection and Prevention System |
| IDS | Intrusion Detection System |
| IETF | Internet Engineering Task Force |
| IoT | Internet of Things |
| IoT | Internet of Things |
| IPS | Intrusion Prevention System |
| LDAP | Lightweight Directory Access Protocol |
| NFSHUNT | Firewall based on Linux's Netfilter |
| NIDS | Network Intrusion Detection System |
| NIPS | Network Intrusion Prevention System |
| NSM | Network Security Management |
| PDoS | Permanent Denial of Service |
| PIN | Personal Identification Number |
| PKI | Public Key Infraestructure |
| QC | Quantum Cryptography |
| QKD | Quantum Key Distribution |
| RSA | Rivest, Shamir and Aldeman Algorithm |
| SDN | Software Defined Networking |
| SSH | Secure Shell |
| SSL | Secure Socket Layer |
| VC | Visual Chyptography |
| WWII | Second World War |
| XOR | Or Exclusice Operation |